\documentclass[12pt]{article}
\usepackage{epsfig,graphicx}

\title{Spin interactions in mesons in strong magnetic field}
\author{  Yu.A.Simonov,\\ Institute of Theoretical and Experimental
Physics\\ 117118, Moscow, B.Cheremushkinskaya 25, Russia}
\date{}

\newcommand{\be}{\begin{equation}}
\newcommand{\ee}{\end{equation}}

\def\la{\mathrel{\mathpalette\fun <}}
\def\ga{\mathrel{\mathpalette\fun >}}
\def\fun#1#2{\lower3.6pt\vbox{\baselineskip0pt\lineskip.9pt
\ialign{$\mathsurround=0pt#1\hfil ##\hfil$\crcr#2\crcr\sim\crcr}}}

\newcommand{{\SD}}{\rm SD}

\newcommand{\vex}{\mbox{\boldmath${\rm x}$}}
\newcommand{\vey}{\mbox{\boldmath${\rm y}$}}
\newcommand{\ver}{\mbox{\boldmath${\rm r}$}}
\newcommand{\vesig}{\mbox{\boldmath${\rm \sigma}$}}

\newcommand{\vep}{\mbox{\boldmath${\rm p}$}}

\newcommand{\vez}{\mbox{\boldmath${\rm z}$}}

\newcommand{\veA}{\mbox{\boldmath${\rm A}$}}

\newcommand{\ven}{\mbox{\boldmath${\rm n}$}}
\newcommand{\veu}{\mbox{\boldmath${\rm u}$}}

\newcommand{\veta}{\mbox{\boldmath${\rm \eta}$}}
\newcommand{\veB}{\mbox{\boldmath${\rm B}$}}
\newcommand{\veH}{\mbox{\boldmath${\rm H}$}}
\newcommand{\veE}{\mbox{\boldmath${\rm E}$}}

\newcommand{\vegam}{\mbox{\boldmath${\rm \gamma}$}}

\newcommand{{\Mc}}{\mathcal{M}}

\newcommand{\lan}{\langle}
\newcommand{\ran}{\rangle}

\begin{document}

\maketitle
\begin{abstract}
Spin interactions in relativistic quark-antiquark system in magnetic field is
considered in the framework of the relativistic Hamiltonian, derived from the
QCD path integral. The formalism allows to separate spin-dependent terms from
the basic spin-independent interaction, contained in the Wilson loop, and
producing confining and gluon exchange interaction. As a result one obtains
relativistic spin-spin interaction $V_{ss}$, generalizing its nonrelativistic
analog. It is shown, that in  large magnetic field $eB, V_{ss}$ modifies and
produces hyperfine shifts which grow linearly with $eB$ and preclude the use of
perturbation theory. We also show, that tensor forces for $eB \neq 0$ are
active in all meson states, but do not grow with $eB$

\end{abstract}

\section{Introduction}
 The spin-dependent terms in the interaction of two fermions in QED have a long
 history \cite{1,2,3,4}, where spin terms were deduced based on Pauli or  Dirac
 form  of Hamiltonians.  In particular the hyperfine (hf) interaction term was
 introduced by Fermi in \cite{3} in the nonrelativistic form. Later on the  QED
 perturbation theory was used to consider hf interaction in higher orders
 \cite{6} see \cite{7} for a review, and also in presence of magnetic field
 (m.f.) \cite{8}.

 An important tool for the higher order relativistic treatment of bound states
 in QED was Bethe-Salpeter equation  (BSE) \cite{9}, or its instantaneous
 Salpeter form \cite{10}.

 It is the purpose  of the present paper  to consider spin-dependent effects in the dynamics of the quark-antiquark $(q\bar q)$ mesons in
strong m.f. The physical interest of such systems is due to high magnetic
fields, which are expected in neutron stars, early universe and heavy-ion
collisions, see \cite{10*} for discussion and references.

In such  QCD systems, as mesons or baryons, the main part of dynamics is of
nonperturbative (np) origin, e.g. the confinement interaction $V_{\rm conf}$,
which cannot be  described by perturbative  propagator particle exchanges. In
addition the relative time problem existing in BSE is more fundamental in QCD,
where the vacuum field correlations  with period $\lambda$, yielding
confinement (and all strong interactions, including spin-dependent ones) are
shorter in time, than physical particle exchanges, so that each quark in meson
propagates in the already time-averaged vacuum \cite{11}. Therefore to
establish the fine and hf structure of strong interactions, one must use
instead of BSE another relativistic formalism, namely the relativistic path
integrals method, based on the so-called Fock-Feynman-Schwinger Representation
\cite{12, 13}. Recently a new form of this method was derived in \cite{14},
where a new integral representation of the $q\bar q$ Green's function was
given.

Basically, the main part of the strong dynamics (non spin-dependent) is given
by the Wilson loop average and the methods, based on FFSR, give all meson,
baryon, glueball and hybrid  properties in good agreement with experimental
data, see \cite{15}, \cite{16} for reviews.

Of special importance for us here are spin-dependent forces in QCD (which we
shall test applying high m.f.). The latter have been derived for large  quark
masses in \cite{17}, and in the framework of our method for any masses  in
\cite{11,18,18*, 18**,18***, 18iv}.

The most important development in the study of the spin-dependent (SD)
interactions consists of three advancements: \begin{enumerate}
    \item  Definition of perturbative and np, SD contributions and expressions
    for them in terms of the standard field correlators.
    This was done  for large masses in  \cite{11, 18}.

    \item Derivation of SD terms in the relativistic systems for low and zero
    mass quarks, done in  \cite{15,18*} and \cite{18**}.
    \item Definition of SD terms for nonzero temperature and in the deconfined
    QCD phase, which was done in \cite{18***}.
    \item Reexamining of the SD forces and suppression of spin-orbit forces in
    hadrons by the string motion, \cite{18iv}.
\end{enumerate}

In the present paper we perform calculation of SD forces in the relativistic
$q\bar q$ system in the presence of   m.f. $eB$, which strongly modifies these
forces, and in particular yields a linear growth in $eB$ for the hf term. We
also show, that m.f. creates deformation of the meson shape and thus stipulates
tensor forces in the originally spherical meson.

A possible contradiction of these effects with the positivity property of the
Green's function is formulated and possible outcomes are discussed.

 In the same formalism we calculate the selfenergy correction to the meson mass in
strong m.f.

The paper is organized as follows: in section 2 we define spectral properties
of Green's function in m.f., in section 3 we derive the $q\bar q$ Green's
function and relativistic Hamiltonians in m.f. from FFSR, in section  4 we
concentrate on SD forces in m.f. and write down explicit expressions for them.
The section 5 is devoted to tensor forces in m.f. The last section contains
discussion of the results and an outlook.  One  appendix contains details of
derivations.

\section{Absence of pair creation in color Euclidean and magnetic fields}

Consider quark-antiquark system in the external magnetic field, interacting
with color vacuum Euclidean and perturbative fields, so that the covariant
derivative is \be D_\mu  \equiv D_\mu (A_\mu^{(e)}, ~ A_\mu^a t^a) =
\partial_\mu - ie A_\mu^{(e)} (x)  - ig A^a_\mu t^a, \label{1}\ee
where $A_\mu^{(e)}$ can be decomposed into the magnetic (``Euclidean'') part
$A_\mu^{(B)}$ and the electric part, e.g.

$$A^{(e)}_\mu = A_\mu^{(B)}+A_\mu^{(e)}{ \rm (pert)},$$

 \be \veA^{(B)} = \frac12 (\vex\times \veB), ~~ A_4^{(B)}=0\label{2}\ee for
the constant magnetic field $\veB$ along $z$ axis. Then the partition function
averaged over nonperturbative  vacuum can be written as \be \lan Z\ran_A = \lan
\int D \bar \psi D\psi \exp (- i \bar \psi (m+ \hat D )\psi)\ran_A,\label{3}\ee
where  the averaging over gluonic fields is implied, \be\lan K\ran_A = \int
K\exp \left(-\frac14 \int (F^a_{\mu\nu} F^a_{\mu\nu}) d^4 x\right)
DA,\label{4}\ee and we disregard for simplicity   gauge fixing and ghost terms.

Integration over $D\bar \psi D\psi$ in(\ref{3}) yields the standard answer
(where the proper renormalization is implied), \be \lan Z\ran_A = \left\lan
\exp \frac12 tr \ln (m^2-\hat D^2)\right\ran_A=\left\lan\exp \left(\frac12 tr
\int^\infty_0 \frac{ds}{s} e^{-s  (m^2-\hat
D^2)}\right)\right\ran_A.\label{5}\ee

Now the question of stability of the vacuum in the given external fields can be
associated with the  nonnegativity of the operator $(m^2- \hat D^2)$, since
otherwise negative eigenvalues of this operator would provide imaginary part in
the exponent of (\ref{5}), which implies finite probability of pair creation,
as it is clearly seen in the Schwinger expression for the pair creation in  the
constant electric (nonEuclidean)  field \cite{20}.

In what follows we show that in purely Euclidean (colorelectric  or
colormagnetic) fields  and in  the magnetic field (external and perturbative)
the operator $(m^2 - \hat D^2)$ is nonnegative and hence pair creation is
absent and vacuum stability is ensured.

To this end  consider the Euclidean operator \be i D_\mu \gamma_\mu =
\gamma_\mu (i \partial_\mu + e A^{(e)}_\mu(x) + g A^a_\mu t^a),\label{6}\ee
with Euclidean and hermitian $\gamma$ matrices, $\gamma_4 \equiv \beta;
\gamma_i = - i\beta \alpha_i; \gamma_i^+ = \gamma_i,$ and with hermitian
$A_i^{(e)}$ and $A_4 \equiv A_4^a t^a, A^+_4 = A_4,$ while $A_4^{(e)} \equiv
0$, so that \be (iD_\mu \gamma_\mu)^+ = i D_\mu \gamma_\mu,\label{7}\ee and for
the eigenfunctions $u_n$ and eigenvalues $\lambda_n$ we have \be i D_\mu
\gamma_\mu u_n = \lambda_n u_n, ~~ \lambda_n ~{\rm real}.\label{8}\ee Hence \be
- \hat D^2 u_n = (i D_\mu\gamma_\mu)^2 u_n = \lambda^2_n u_n, ~~\lambda^2_n
\geq 0,\label{9}\ee and $\parallel m^2 - \hat D^2\parallel\geq m^2$, which
implies vacuum stability and no pair creation for any $m\geq 0$.

 In terms of the quark propagator $S= \frac{i}{\hat D +m }$, one has
 representations
 \be iS = \sum_n \frac{u_n (x) u^+_n (y)}{\lambda_n - im} , ~~ \left(
 \frac{1}{m^2-\hat D^2}\right)_{xy} = \sum\frac{u_n (x) u^+_n
 (y)}{m^2+\lambda_n^2}.\label{10a}\ee
Writing  $u_n = \left(\begin{array}{l} \varphi_n\\\chi_n\end{array}\right)$,
  one obtains the following equations $$  -\tilde D^2 +\vesig ( g\veE - g\veH - e\veB)
 (\varphi_n - \chi_n) = \lambda^2_n ( \varphi_n - \chi_n),$$
\be -\tilde D^2- \vesig ( g\veE + g\veH +e\veB)
 (\varphi_n + \chi_n) = \lambda^2_n ( \varphi_n+ \chi_n),\label{11a}\ee
 where $\tilde D^2= (\partial_\mu -ie A_\mu^{(e)} -ig A_\mu)^2$, and $A_\mu,
 \veE, \veH$ correspond to the color fields e.g. $A_\mu \equiv A_\mu^a
 t^a_{\alpha\beta},$ while $A_\mu^{(e)}$ is the electromagnetic field, and
 $\veB$ is the external magnetic field. From the system (\ref{11a}) one can
 see, that for hermitian Euclidean fields $A_\mu^{(e)}, A_\mu, \veE, \veH,
 \veB$ the eigenvalues $\lambda^2_n$ are real and the Green's function
 $\frac{1}{m^2-\hat D^2}$ has only positive  set of eigenvalues.

Note, that in real electric field $A_0^{(e)}, A_4^{(e)} = i A_0^{(e)}$ and the
property of nonnegativity of $(- \hat D^2)$ is violated, implying possible
quark pair creation, while in absence of $A_0^{(e)}$, but with real vacuum
color field $A_4^+ = A_4$ and hence real colorelectric vacuum field $E^{\rm
vac}_i \sim
\partial_i A_4$, vacuum is stable with known vacuum condensate $((E^{\rm
vac}_i)^2 +(H^{\rm vac}_i)^2)$. However, the quasizero modes with small
$\lambda_n$ can accumulate, implying Chiral Symmetry Breaking (CSB), signalled
by the Banks-Casher formula  \cite{21}, see \cite{22} for   review and
discussion of CSB from this point of view. As we shall see, perturbative
colormagnetic interactions in the lowest order, in the external magnetic field
can violate positivity condition (9), signalling the divergence of the
perturbative series.

\section{Relativistic $q\bar q$ Green's function in magnetic field in the path-integral  form}

The quark Green's function in magnetic field can be written as \be S_q (x,y) =
(m+ \hat D)^{-1}_{xy} = (m-\hat D)_x (m^2- \hat D^2)^{-1}_{xy},\label{10}\ee
where \be m^2-\hat D^2 =m^2- D^2_\mu - g \sigma_{\mu\nu} F_{\mu\nu} - e
\sigma_{\mu\nu} F^{(e)}_{\mu\nu}\label{11}\ee and $D_\mu$ is given in
(\ref{6}), while \be \sigma_{\mu\nu} F_{\mu\nu} = \left( \begin{array}{ll}
\vesig \veH&\vesig \veE\\\vesig\veE& \vesig \veH\end{array}\right),  ~ ~
\sigma_{\mu\nu} F_{\mu\nu}^{(e)} = \left( \begin{array}{ll} \vesig \veB&0
\\0& \vesig \veB\end{array}\right)\label{12}\ee

One can use for $S_q$ the path-integral form \cite{12,14} \be S_q (x,y) = (m-
\hat D)_x \int^\infty_0 ds D^4z \Phi_z^{(F)} (x,y) e^{- K  }\label{13}\ee where
\be \Phi_z^{(F)} (x,y) = P_A \exp \left(ie \int^x_y A_i^{(e)} dz_i + ig
\int^x_y A_\mu dz_\mu+e \sigma_{\mu\nu} \int^s_0  F_{\mu\nu}^{(e)} d\tau+g
\sigma_{\mu\nu} \int^s_0 F_{\mu\nu} d\tau\right) ,\label{14}\ee

\be K = \int^s_0 \left[ m^2+\frac14 \left( \frac{dz_\mu}{d\tau}\right)^2\right]
d\tau .\label{15}\ee

Then the  $q\bar q$  Green's  function can be written as
\be  G_{q_1\bar q_2} (x,y) = \int^\infty_0 ds_1 \int^\infty_0 ds_2
(D^4z^{(1)})_{xy} (D^4z^{(2)})_{xy} \lan  \hat TW_\sigma (A)\ran_AW_\sigma
(A^{(e)})\ran , \label{18aa}\ee where \be W_\sigma (A) = P\exp (ig \int_C A_\mu
(z) dz_\mu + g\int^{s_1}_0 \sigma_{\mu\nu} F_{\mu\nu} (\tau_1) d\tau_1 - g
\int^{s_2}_0 \sigma_{\mu\nu} F_{\mu\nu} (\tau_2) d\tau_2),\label{19aa}\ee

 \be W_\sigma( A^{(e)})=  \exp (ie_1 \int^x_y A^{(e)}_\mu
dz^{(1)}_\mu -ie_2 \int^x_y A^{(e)}_\mu dz^{(2)}_\mu +e_1\int^{s_1}_0 d\tau_1
(\sigma_{\mu\nu} F_{\mu\nu}^{(e)}) -e_2\int^{s_2}_0 d\tau_2 (\sigma_{\mu\nu}
F_{\mu\nu}^{(e)})),\label{12a}\ee

\be \hat T = \frac14 tr (\Gamma_1 (m_1 -\hat D_1) \Gamma_2 (m_2 -\hat D_2))\exp
(-K_1-K_2),\label{13a}\ee and $\Gamma_1 = \gamma_{\mu},\ \Gamma_2 =
\gamma_{\nu}$ for vector currents, while $\Gamma_1 =\Gamma_2 = \gamma_5$ for
pseudoscalars, and the symbol $tr$ implies summation over color and Dirac
indices and refers to all terms.

At this point we introduce  new variables  $\omega_i$  in the path integral
(\ref{13}), defined via the connection between the proper time $\tau_i$ and the
real Euclidean time $t_i^E = z_4 (\tau_i)$ (see details in Appendix 1 of
\cite{14}), and integrating over time fluctuations using

 $$ s_i =
\frac{T}{2\omega_i}, ~~ ds_i =-\frac{T d \omega_i}{2\omega^2_i;} ~~ d\tau_i =
\frac{dt^E_i}{2\omega_i},$$~~

\be  \int^\infty_0 ds_i (D^{4}z^{(i)})\Phi_z({x,y}) e^{-K}  =
 {{T} \int^\infty_0 \frac{ d\omega_i  }{2 \omega_i^{2}}(D^{3}
 z^{(i)})_{\vex\vey}e^{-K(\omega_i)}\lan \Phi_z({x,y})\ran_{\Delta z_4}}
  .\label{15a}\ee
 where $\lan ~\ran_{\Delta z_4}$ means the averaging over time fluctuations,
 which can be written in terms of the averaged Wilson line \cite{14}

\be \lan \Phi_z(x,y)\ran_{\Delta z_4} = \sqrt{\frac{\omega_i }{2\pi T}}
\Phi_z(x,y),\label{q23}\ee and $K(\omega)$ is \be  K(\omega) =\int^T_0 dt_E
\left(\frac{\omega}{2} + \frac{m^2}{2\omega} + \frac{\omega}{2}
\left(\frac{d\vez}{dt_E}\right)^2\right)\label{q24}\ee

 $$ T= |x_4 - y_4|.$$
 In this way the
path integral in $ ds_i D^4z^{(i)}$ is replaced by $d\omega_i(D^3z^{(i)})$, and
the $q\bar q$ Green's function (\ref{18aa}) can be written as

 \be G_{q_1\bar q_2}
(x,y) =  \frac{T}{2\pi} \int^\infty_0\frac{ d\omega_1  }
{\omega_1^{3/2}}\int^\infty_0\frac{ d\omega_2  }{ \omega_2^{3/2}} (D^{3}
z^{(1)}D^{3} z^{(2)} )_{\vex\vey}\overline{\lan\hat TW_\sigma (A) W_\sigma
(A^{(e)})\ran},\label{36a}\ee where \be \overline{\lan\hat TW\ran } = 4 tr
Y\overline{\lan W W\ran} \exp (-K(\omega_1) - K(\omega_2))\label{q26}\ee
$\overline{W} $ is the averaged over time fluctuations Wilson loop, see
\cite{14}, and $Y=\frac14 \Gamma_1 (m_1 - i \hat P_1) \Gamma_2 (m_2 - i\hat
p_2)$. Note, that the operator ordering is not taken into account in
(\ref{q26}) and we shall take care of it below  computing SD terms.

Using cluster expansion and nonabelian Stokes theorem  \cite{23} one can
rewrite $\overline{W_\sigma}$ as

\begin{eqnarray}
\langle{\rm Tr}W(C)\rangle&=&\langle{\rm Tr}\exp ig\int d\pi_{\mu\nu}(z)F_{\mu\nu}(z)\rangle\nonumber\\[-2mm]
\label{W1}\\[-2mm]
=\exp\sum^\infty_{n=1}\frac{(ig)^n}{n!}&\displaystyle\int& d\pi(1) \ldots\int d
\pi(n)\langle\langle F(1)\ldots F(n)\rangle\rangle\nonumber,
\end{eqnarray}
where $d\pi_{\mu\nu} \equiv ds_{\mu\nu} + \sigma_{\mu\nu}^{(1)} d{\tau_1} -
\sigma_{\mu\nu}^{(2)} d\tau_2,$ and $ds_{\mu\nu}$ is an area element of the
minimal surface, which can be constructed using straight lines, connecting the
points $z_\mu^{(1)} (t)$ and $z_\nu^{(2)} (t)$ on the paths of $q_1$ and $\bar
q_2$ at the  same time $t$ \cite{11,23a}. Note, that $z_4^{(1)}(t) =
z_4^{(2)}(t) =t$.  Then the spin-independent part of the exponent reduces to
the confinement term $V_{\rm conf}(r)$ plus color Coulomb potential $V_{\rm
Coul}$, while spin-dependent part $V_{SD}$ depends also on proper time
variables $\tau_1, \tau_2$, (see \cite{11,18} for derivation and discussion).
For the case of zero quark orbital momenta with the minimal surface, discussed
above, one obtains a simple answer for $\langle W_{\sigma}(A)\rangle_{A}$,
which we shall derive below.

 The average
$\langle\langle\ldots\rangle\rangle$ stands for connected correlators, for
example, for the bilocal correlator, $\langle\langle
F(1)F(2)\rangle\rangle=\langle F(1)F(2)\rangle-\langle F(1)\rangle\langle
F(2)\rangle$, and $F_{\mu\nu}=\partial_\mu A_\nu-\partial_\nu A_\mu- ig [A_\mu,
A_\nu]$ is the vacuum field strength. Obviously, due to the $O(4)$ rotational
invariance and colour neutrality of the vacuum, $\langle\langle
F\rangle\rangle=\langle F\rangle=0$.

In the Gaussian approximation for the vacuum, when only the lowest, bilocal
correlator is retained, one has, with the accuracy of a few per cent (see
Ref.~\cite{18**,18iv} for the discussion): \be \langle Tr
W(C)\rangle\propto\exp\left[-\frac12\int_Sd\pi_{\mu\nu}(x)d\pi_{\lambda\rho}(x')
D_{\mu\nu\lambda\rho}(x-x')\right], \label{W2} \ee where \be
D_{\mu\nu\lambda\rho}(x-x')\equiv\frac{g^2}{N_c}\langle\langle{\rm
Tr}F_{\mu\nu}(x)\Phi(x,x')F_{\lambda\rho}(x')\Phi(x',x)\rangle\rangle.
\label{24a}\ee This bilocal correlator of gluonic fields can be expressed
through only two gauge-invariant scalar functions $D(u)$ and $D_1(u)$ as
\cite{25} \be
D_{\mu\nu\lambda\rho}(u)=(\delta_{\mu\lambda}\delta_{\nu\rho}-\delta_{\mu\rho}\delta_{\nu\lambda})D(u)
+\frac12\left[\frac{\partial}{\partial
u_\mu}(u_\lambda\delta_{\nu\rho}-u_\rho\delta_{\lambda\nu})+\left(
\begin{array}{l}
{\mu\leftrightarrow\nu}\\{\lambda\leftrightarrow\rho}\end{array}\right)\right]D_1(u).
\label{Dcordef} \ee The correlator $D(u)=D(u_0,|\veu|)$ contains   a
nonperturbative part and it is responsible for confinement:  the QCD string
formation at large interquark separations.  Indeed $\int ds_{i4} (u) \int
ds_{i4}(v) D(u-v)= S \int d^2 (u-v) D(u-v),$ where $S$ is the total area
between the averaged quark trajectories. The fundamental string tension can be
calculated from the area law asymptotics of (\ref{W2}) for large area loop and
is expressed as a double integral: \be \sigma=\int\int d^2 (u-v)
D(u-v)=2\int_0^\infty d\nu\int_0^\infty d\lambda D(\nu,\lambda). \label{sigma}
\ee




Using (\ref{19aa}) with $\sigma_{\mu\nu} F_{\mu\nu}\equiv 0$, one obtains
spin-independent terms in the $q\bar q$ interaction \cite{25}. \be V_0 (r) =
V_{conf} (r) + V_{OGE} (r), \label{27a}\ee \be V_{conf} (r) = 2r \int^r_0
d\lambda \int^\infty_0  d\nu D(\lambda, \nu) \to \sigma r, (r\to
\infty)\label{28a}\ee \be V_{OGE} = \int^r_0 \lambda d\lambda \int^\infty_0
d\nu D_1^{ pert} (\lambda, \nu) =- \frac43 \frac{\alpha_s}{r},\label{29a}\ee
 where we keep only perturbative part of $D_1$ and to the lowest order
 $D_1^{pert} (\lambda, \nu) = \frac{16 \alpha_s}{3 \pi (\lambda^2 +\nu^2)^2}$.
 We now turn to the spin-dependent terms of the  $q\bar q$ interaction,
 $V_{SD}$, and we shall be interested only in the zero orbital moment states
 for simplicity, hence only spin-spin interaction term $V_{ss}$ will be treated
 below.

The spin--dependent terms in the interquark interaction are generated by the
combination $\sigma_{\mu\nu}F_{\mu\nu} $ present in Eq.(\ref{12}) and therefore
one needs correlators of the colour-electric and colour-magnetic fields, as
well as mixed terms, separately. They immediately follow from the general
expression (\ref{Dcordef}) and read \cite{26}: $$ \frac{g^2}{N_c}\lan\lan {\rm
Tr} E_i(x)\Phi
E_j(y)\Phi^\dagger\ran\ran=\delta_{ij}\left(D^E(u)+D_1^E(u)+u^2_4\frac{\partial
D_1^E}{\partial u^2}\right)+ u_iu_j\frac{\partial D_1^E}{\partial u^2},$$
$$\frac{g^2}{N_c}\lan\lan {\rm Tr} H_i(x)\Phi
H_j(y)\Phi^\dagger\ran\ran=\delta_{ij}\left(D^H(u)+D_1^H(u)+\veu^2\frac{\partial
D_1^H}{\partial\veu^2}\right)- u_iu_j\frac{\partial D_1^H}{\partial
u^2},\label{Hs0}$$ \be\frac{g^2}{N_c}\lan\lan {\rm Tr} H_i(x)\Phi
E_j(y)\Phi^\dagger\ran\ran=\varepsilon_{ijk} u_4u_k\frac{\partial
D_1^{EH}}{\partial u^2}, \label{HE0} \ee  where $u_\mu=x_\mu-y_\mu$, $u^2=u_\mu
u_\mu$. We keep here the superscripts $E$ and $H$ in the correlators $D$ and
$D_1$ in order to distinguish  in principle the electric and magnetic parts of
the correlators and thus to be able to consider a nonzero temperature $T$ and
to distinguish Euclidean and Minkowskian contributions. Indeed, while $D^E=D^H$
and $D_1^E=D_1^H$ at $T=0$, at higher temperatures they behave differently. In
particular, above the deconfinement temperature, $T>T_c$, the electric
correlator $D^E$ disappears, whereas $D_1^E$ and the magnetic correlators
survive.

It is clear from (\ref{12}), that in the norelativistic limit, when in
(\ref{15}) in $\int \sigma_{\mu\nu} F_{\mu\nu }{d \tau} $, $d\tau_i =
\frac{dt_i}{2m_i}$, only the upper left corner of (\ref{12}), i.e.
$(\vesig\veH)$, will contribute to $V_{ss}$. The corresponding derivation was
done in \cite{11,18} and gives \be V_{ss} (r) = \frac{\vesig_1 \vesig_2}{12
m_1m_2} V_4 (r) + \frac{3(\vesig_1 \ver) (\vesig_2\ver) - \vesig_1 \vesig_2
r^2}{12m_1 m_2 r^2} V_3(r)\label{32a}\ee

where 
\be  V_3(r) =- \int^\infty_{-\infty} d\nu r^2 \frac{\partial D_1^{pert} (r,
\nu)}{\partial r^2} = \frac{4\alpha_s}{r^3},\label{33a} \ee

\be  V_4(r) = \int^\infty_{-\infty} d\nu \left(3D_1^{pert}(r,\nu)+2  r^2
\frac{\partial D_1^{pert} (r, \nu)}{\partial r^2}\right) = \frac{32\pi
\alpha_s}{ 3}\delta^{(3)} (\ver),\label{34a} \ee

It is our purpose below to calculate $V_{ss}$ in the relativistic $q\bar q$
system and in arbitrarily large magnetic field $B$, and to this end we shall
use below first the path integral formalism \cite{13,14}, deriving the general
structure of   the $q\bar q$ Green's  function, and in the next section we
shall determine how  relativistic $V_{ss}$ expressions depend on $B$.

First we need to find the the Euclidean action $S^E_{q_1\bar q_2}$ in terms of
$\omega_1, \omega_2$ and common time $t^E$ of the $q\bar q$ system  at $t_1^E =
t_2^E = t^E$. To this end we define the Euclidean Lagrangian $L^E_{q_1 \bar
q_2} $. We write $\frac{d z^{(i)}}{d\tau_i} = 2\omega_i\frac{dz^{(i)}_k}{dt^E}
= 2\omega_i \dot{z}_k,\ k = 1,2,3$. Then all terms in the exponents in
(\ref{19aa}), (\ref{12a}) and (\ref{13a}) can be represented as $\exp(-\int
dt^E L_{q_1 \bar q_2}^E)$ and thus we arrive at the following action $$
S^E_{q_1\bar q_2} = \int^{T_E}_0 dt^E \left[ \frac{\omega_1+\omega_2}{2} +
 \sum_i \left( \frac{\omega_i}{2} (\dot z_k^{(i)})^2 \right)-\right.$$ \be- ie_i A_k^{(e)} \dot z^{(i)}_k + \frac{m_1^2}{2\omega_1} +
 \frac{m_2^2}{2\omega_2}+ e_1 \frac{\vesig_1 \veB}{2\omega_1}+e_2 \frac{\vesig_2
 \veB}{2\omega_2}
  \left.+\sigma |\vez^{(1)} -\vez^{(2)}| -\frac43
 \frac{\alpha_s}{|\vez^{(1)}-\vez^{(2)}| }\right]+ S^E_F,\label{16a}\ee
where $S^E_F$ contains  $(\sigma F)$ terms.
 Here $A_k^{(e)}$ is the $k$--th
 component of the QED vector potential, $\sigma$ is the QCD string tension and the
 contribution of terms $(\sigma_1 F), (\sigma_2F)$ is separated in $S^E_F$.
  The next step
 is the transition to the Minkowski metric and the construction of the Hamiltonian. This is easy, since confinement is already expressed
 in terms of  the string tension. We have $\exp(-\int L^E dt_E) \rightarrow \exp(i \int L^M dt_M),\ t_E \rightarrow it_M$,
  and \be H_{q_1\bar q_2} = \sum_i \dot z_k^{(i)} p^{(i)}_k -L_M, ~~ p^{(i)}_k =
 \frac{\partial L^M}{\partial\dot z_k^{(i)}} =\omega_i \dot z_k^{(i)} + e_i
 A_k^{(e)}.\label{17a}\ee

As a result one obtains (back in the Euclidean time $T=|x_4-y_4|$),

\be G(x,y) = \frac{T}{2\pi}\int^\infty_0\frac{ d\omega_1  }{
\omega_1^{3/2}}\int^\infty_0\frac{ d\omega_2  }{ \omega_2^{3/2}}  4 tr  Y \lan
\vex |e^{- H_{q_1 \bar q_2} T}|\vey\ran, \label{39q}\ee

 \be H_{q, \bar q_2} =H_0 + H_\sigma +W,\label{1a}\ee
 \be W=V_{\rm conf}+V_{OGE}+\Delta M_{SE}+\Delta M_{ss}, \label{4a}\ee
 where
\be H_0 =\sum^2_{i=1} \frac{\left(\vep^{(i)} - \frac{e_i}{2} (\veB\times
\vez^{(i)})\right)^2+ m^2_i +\omega^2_i}{2\omega_i},\label{2a}\ee

 \be H_\sigma
=- \frac{e_1\vesig_1\veB}{2\omega_1}
-\frac{e_2\vesig_2\veB}{2\omega_2}.\label{3aa}\ee

Here the terms $\Delta M_{SE} $ and $\Delta M_{ss}$ are produced by $S^E_F$,
and we shall find them as a first order correction. But before that we must
treat the $\omega_i$ dependence either in the path integral (\ref{36a}), or the
Hamiltonian (\ref{1a}). In the path integral $\omega_i$ play the role of quark
 energy  parameters, and one can  use  the spectral decomposition in (\ref{39q}) to rewrite it as

\be G(x,y) = \frac{T}{2\pi}\int^\infty_0\frac{ d\omega_1  }
{\omega_1^{3/2}}\int^\infty_0\frac{ d\omega_2  }{ \omega_2^{3/2}}
\sum^\infty_{n=0} 4 tr  Y \lan \vex |n\ran e^{- M_n  T}\lan n|\vey\ran,
\label{40q}\ee
 At large $T$ one can  use the stationary point method,
 and one defines $  \omega_i^{(0)} $ from
the extremum values of $M_n (\omega_1, \omega_2)$, namely for the Hamiltonian
$\bar H$,

\be H_0+H_\sigma +V_{conf} +V_{OGE} = \bar H;~~  \bar
H\Psi=M_n^{(0)}\Psi,\label{5a}\ee and $\omega_i^{(0)}$ is defined  from the
condition
 \be \frac {\partial
M_n^{(0)}(\omega_1,\omega_2)}{\partial\omega_i}|_{\omega_i=\omega_i^{(0)}}=0,~~
i=1,2.\label{6a}\ee

To have an idea of the possible meson masses and the values of
$\omega_i^{(0)}$, which we shall use below, it is instructive to consider as in
\cite{11} the main part of the Hamiltonian, i.e. \be \tilde H = H_0 +  H_\sigma
+ V_{\rm conf}, ~~ \tilde H \tilde \psi = \tilde M \tilde \psi \label{q49}\ee
and replace $V_{\rm conf}$ by  the quadratic term, \be V_{\rm  conf} (r) =
\sigma  r \to \tilde V_{\rm conf} (r) = \frac{\sigma}{2}  \left(
\frac{r^2}{\gamma} + \gamma\right).\label{q50}\ee where $\gamma$ is the
variational parameter, yielding some 5\% accuracy in the replacement
(\ref{q50}). Then the resulting mass $\tilde M$ can be found explicitly as \be
\tilde M (\omega_1, \omega_2, \gamma) = \varepsilon_{n_\bot , n_z} +
\frac{m_1^2+\omega^2_1 - e_1\veB \vesig_z}{2\omega_1} +\frac{m_2^2+\omega^2_2 -
e_2\veB \vesig_z}{2\omega_2},\label{q51}\ee where \be \varepsilon_{n_\bot, n_z}
= \frac{1}{2\tilde \omega} \left[ \sqrt{ e^2 B^2 + \frac{4\sigma\tilde
\omega}{\gamma}} (2n_\bot +1)+ \sqrt{\frac{4\sigma \tilde
\omega}{\gamma}}\left(n_z + \frac12\right)\right] + \frac{\gamma \sigma}{2},
\label{q52}\ee As a result one can estimate the masses $\tilde M$ and
$\omega_i^{(0)}$ at large m.f., since the basic pattern is defined by relative
signs  of $eB$ terms in $\varepsilon_{n_\bot, n_z}$ and $H_\sigma$.

Indeed, for $eB\gg \sigma$ one can write $\tilde M  \cong \sum_{i=1,2}
\frac{m^2_i + \omega^2_ i+ | e_i B| -e_i\vesig_i \veB}{2\omega_i}$ and
\be\omega_i^{(0)} \approx \Omega_i\equiv\sqrt{m^2_i + |e_i B| - e_i \vesig_i
\veB} +O(\sqrt{\sigma}),~~\tilde M \cong \Omega_1+\Omega_2.\label{52*}\ee

Thus, for the neutral meson with $e_2 = - e_1$, and $\sigma_{1z}, \sigma_{2z} =
(++), \omega_{++}$ is growing as $\sqrt{e_1 B}$, while for the $(+-)$ state
$\omega_{+-}$ is tending to a constant. One can also find the wave function

\be \tilde \psi(\veta) = \frac{1}{\sqrt{\pi^{3/2} r^2_\bot r_0}} \exp \left(
-\frac{\eta^2_\bot}{2r^2_\bot} -
 \frac{\eta^2_{z}}{2r^2_0}\right),~ \veta= \vez^{(1)} - \vez^{(2)},\label{q53}\ee and
  \be r_\bot =  {\frac{2}{eB}} \left( 1+ \frac{4\sigma\tilde
 \omega}{\gamma_0 e^2B^2}\right)^{-1/2},~ r_0 = \left( \frac{\gamma}{\sigma
 \tilde \omega}\right)^{1/4}.\label{q54}\ee

 At large $eB\gg \sigma$, one has $r^2_\bot \approx  \frac{2}{eB}, r_0 \approx
  const \approx \frac{1}{\sqrt{\sigma}},$ and hence
 \be |\tilde \psi (0)|^2 \approx \frac{eB\cdot \sqrt{\sigma}}{2\pi^{3/2}},~~
 (eB\gg \sigma).\label{q55}\ee

 This is the focussing effect of m.f., which is most important in SD forces, as
 well as in other processes \cite{34}.

It is important to stress, that we have kept in $\bar H$ only those terms,
which are the main part of interaction, and therefore in $M_n^{(0)}$ and
$\omega^{(0)}$  they are treated  to all orders, i.e. exactly.  However, the
terms $V_{ss}$ and $\Delta M_{SE}$ are considered only as a perturbation, and
therefore one should substitute there the values $\omega_i^{(0)}$ obtained from
(\ref{5a}), (\ref{6a}), where $V_{ss}$, $\Delta M_{SE}$ do  not enter. The
Hamiltonians $\bar H$ are considered in \cite{14}, and below we shall derive
both $V_{ss}$ and $\Delta M_{SE}$ in the relativistic $q\bar q$ system.

\section{The quark-antiquark spin-dependent interaction in strong magnetic
field}

The advantage of representation (\ref{18aa}),  (\ref{36a}) lies in the fact,
that the only place, where the Dirac $\gamma$ matrices enter, is the local term
$(m-\hat D)$, and it  can be assembled in the factor $Y$ with due  care, while
all the rest nontrivial spin dependence is contained in the $(\sigma F)$ and
$(\sigma B)$ factors (\ref{12}). We shall demonstrate below in this section,
that the correlators $(\sigma^{(i)} F) (\sigma^{(k)} F)$ with $i=k$ define
$\Delta M_{SE}$, while those with $i\neq k $, define $V_{ss}$.

Consider    the Taylor expansion in powers of the color spin interaction
$g\sigma_{\mu\nu} F_{\mu\nu} \equiv g (\sigma F)$, ~ $ m^2 - \hat D^2 = m^2 -
\tilde D^2 - g(\sigma F)$, $\tilde D^2_\mu= D^2_\mu - e(\sigma F^{(e)}), $

$$ \frac{1}{m^2-\hat D^2} = \frac{1}{m^2- \tilde D^2 - g \sigma F} =
\frac{1}{m^2-\tilde D^2} + \frac{1}{m^2-\tilde D^2} g \sigma F
\frac{1}{m^2-\tilde D^2}+$$ \be \frac{1}{m^2-\tilde D^2} g( \sigma
F)\frac{1}{m^2-\tilde D^2} g (\sigma F)\frac{1}{m^2-\tilde
D^2}+...\label{16}\ee

One can define in (\ref{16}) the selfenergy correction to the mass, \be \Delta
m^2 (x,y) = - g(\sigma F)_x \left( \frac{1}{m^2-\tilde D^2} \right)_{xy} g
(\sigma F)_y.\label{17}\ee

\be \bar \Delta m^2 =\int d^4 (x-y) \Delta m^2 (x,y); ~~ \Delta m^2 (x,y) =
-g^2 \sigma_i \sigma_k (\lan H_i H_k\ran +\lan E_iE_k\ran)_{xy} G_0
(x,y)\label{61a}\ee where \be G_0 (x,y) = \left( \frac{1}{m^2-\tilde
D^2}\right)_{xy},\label{62a}\ee while \be \frac{g^2}{N_c} \lan H_i (x) H_k (y)
+ E_i (x) E_k (y)\ran = 2D(x-y)\delta_{ik}. \label{63a}\ee

It was shown  in \cite{27} that in absence of magnetic field and in the limit
of small $m$ and  small vacuum correlation length $\lambda$, one can replace
$\left( \frac{1}{m^2 -\tilde D^2}\right)_{xy}$ by the free propagator
$\frac{1}{(4\pi)^2 (x-y)^2}$, yielding \be\bar \Delta m^2  =
-\frac{3\sigma}{\pi},\label{19}\ee  then the correction (\ref{17}) yields for
the total mass $M^{(0)}_n$ , $\Delta M_n = \sum_i\frac{\bar \Delta
m^2_i}{2\omega_i^{(0)}} $, for zero mass $q$ and $\bar q$ \be M_n^{(0)}
(\omega_0) \to M_n^{(0)} (\omega_0) - \frac{3\sigma}{\pi
\omega_0}.\label{20}\ee
 Note, that $\omega$ plays the role
of the integration variables in (\ref{36a}),  (\ref{39q}) and   is defined from
the stationary point condition (\ref{6a}), where $M^{(0)}_n $ does not include
$\Delta m^2$. However for large $m$ and small $| x-y|\la \lambda$ one should
multiply (\ref{19}) with the coefficient $\eta (m\lambda)<1$ calculated in
\cite{27}.

Consider now the case of constant magnetic field $\veB$ along $z$ axis. One can
calculate the effect  of magnetic field on the selfenergy correction, to this
end  expand \be (\sigma F) \frac{1}{m^2 - D^2_\mu - e\sigma_3 B} (\sigma F) =
(\sigma F) \frac{m^2-D^2_\mu+ e\sigma_3 B}{(m^2-D^2_\mu)^2-(eB)^2}(\sigma F)
\to \frac{\sigma F(G_++G_-)\sigma F}{2} .\label{22}\ee  where $G_{+/-} =
(m^2-D^2_\mu \pm eB)^{-1}_{xy}$, and $m^2_i - D^2_\mu \approx m^2_i + |e_iB|$.
 As was shown in  Eq.(\ref{52*}) for large $eB\gg \sigma$
  one has a large effective mass $\omega_*, \omega_*^2\approx m^2_i +2|e_iB|$ in  one of the Green's functions $G_+, G_-$
   and the corresponding Green's function  will contribute
$\frac{3\sigma}{2\pi \omega_0} \eta (eB)$, where $\eta(eB) \equiv \eta
(\sqrt{2|e_iB |+ m_i^2}\lambda)$  is the coefficient, introduced in \cite{27},
e.g. $\eta(0)=1, \eta(5 ~{\rm GeV})^2)=0.03$, see appendix of \cite{27} for
explicit expression. The final expression for $\Delta M_n $ can therefore be
written as \be  \Delta M_n =- \frac{3\sigma}{2\pi \omega_0} (1+\eta
(eB)),\label{67a}\ee where $\omega_0$ corresponds to the Green's function $G$
of a given quark, when spin terms $(\sigma F)$ are absent.   One can see, that
at large $eB$ the selfenergy correction numerator decreases approximately
twice, as compared to $eB=0,$ $ m=0$, since $\eta (eB\gg \sigma ) \to 0$.

We now turn to the $(q \bar q)$ Green's function and write it in the form \be
\int G_{q_1\bar q_2} (x,y) d^3 (x-y) =  tr \lan [\Gamma_1 (m_1 - i \hat p_1)
e^{-HT} \Gamma_2 (m_2 - i \hat p_2) ]\ran ,\label{68a}\ee where we have taken
into account, that $(\hat \partial -ig \hat A)$ acting in (\ref{68a}) on the
Wilson loop, can be replaced by the momentum operator \cite{28}.

For the case, when $H$ does not contain $\gamma_\mu$ matrices, noncommuting
with $(m_i - i\hat p_i)$, one can rewrite (\ref{68a}) as in  \cite{28}, but now
taking into account spin and isospin nonconservation in m.f., one must keep the
possible eigenvalue dependence on the spin projections,

\be \int G_{q_1 \bar q_2} (x,y) d^3 (x-y) = \sum_{n,\nu} (\varepsilon_r \otimes
\varepsilon_r)_\nu \frac{(M_n^{(\nu)} f^n_\Gamma)^2
e^{-M_n^{(\nu)}T}}{2M_n}\label{69a}\ee with $
\varepsilon_{\gamma_5}=\varepsilon_1 =1, ~~ \varepsilon_V \equiv
\varepsilon_\mu^{(k)}, $ and , and the index $\nu$ denotes a specific
polarization and charge component of quark and antiquark, e.g. $\nu=1$ for
$(\bar u u), <-+|$ component. As a result the quark decay constant of the
$\gamma_\nu$ state is
 \be (f^n_{\Gamma\nu})^2 = \frac{ N_c
\lan Y_{\Gamma\nu} \ran |\psi_n^{(\nu}) (0)|^2}{\bar\omega_1 \bar \omega_2
M_n^{(\nu)}\xi_\nu},\label{70a}\ee where $  \xi_\nu$ occurs due to $\omega_i$
integrations in (\ref{2a}), see appendix of \cite{14}, and   \be \lan
Y_{\Gamma\nu}\ran = \frac14 tr (\Gamma^\nu (m_1 - i \hat p_1) \Gamma^\nu(m_2 -
i \hat p_2)).\label{71}\ee

In the case, when $H$ contain spin-dependent terms, and in addition depends on
magnetic field $B$, one should be more careful with the ordering of operators
$(\sigma F), \vesig \veB$ in $H$ and projectors $(m_1 - i \hat p_1), (m_2-
i\hat p_2)$.

Correspondingly,   $(m-\hat D) $can be rewritten as \be (m_q -\hat D)_x \to
(m_1 - i \hat p_1),~~ (m_{\bar q}- \hat{ \bar D})_x = m_2 - i\hat p_2,
\label{29}\ee where $p_1= (i \omega_1, \vep),~~ p_2 = -(i\omega_2, -\vep)$ and
we take into account, that $D_\mu (x)$ acting on $\Phi_z (x,y)$ yields
$\partial_\mu\to i p_\mu$. \be m_1-i\hat p_1 = m_1 + \omega_1 \gamma_4 - i \vep
\vegam =\left(
\begin{array} {ll} m_1+\omega_1& - i \vesig \vep\\i\vesig \vep &
m_1-\omega_1\end{array}\right)\label{30}\ee

At this point we have  at least two possibilities for the relative ordering of
factors $(m^2_{\bar q} - \hat {\bar D}^2)$ and $(m_{\bar q} - \hat {\bar D})$
in $G_{q\bar q}$.  We shall define this ordering as $$ G_{q\bar q} (x,y) = \lan
tr [ \Gamma_1 S_q (x,y) \Gamma_2 S_{\bar q} (y,x)]\ran_A=$$\be= \lan tr [
\Gamma_1 (m_q-\hat D)_x (m^2_q-\hat D^2)_{xy}^{-1}  \Gamma_2 (m_{\bar
q}-\hat{\bar D})_y (m^2_{\bar q} - \hat{\bar D}^2)_{yx}^{-1}
]\ran_A\label{30a}\ee

We shall  show below, that this ordering yields correct results for spinless
and spin-dependent parts, which can be checked in the nonrelativistic limit,
whereas  other orderings lead to wrong answers. Now

\be m_2-i\hat p_2 = m_2- \omega_2 \gamma_4 - i \vep \vegam =\left(
\begin{array} {ll} m_2-\omega_2& - i \vesig \vep\\i\vesig \vep &
m_2+\omega_2\end{array}\right).\label{31}\ee

If is clear, that in the nonrelativistic situation, $p\ll m$, $ \omega =
\sqrt{p^2 +m^2} \to m $,  the product  $(\Gamma (m_1 - i \hat p_1) \Gamma (m_2-
i \hat p_2))$ tends to the nonrelativistic projector \be \left(\begin{array}
{ll} 4m_1m_2,&0\\0,&0\end{array}\right).\label{35}\ee

We now turn to the spin-dependent terms and make expansion of $m_i^2 - \hat
D^2=m^2_i-D^2_\mu-e\sigma B-g(\sigma F)$ in powers of $g(\sigma F)$: \be (m_1^2
- \hat D^2)^{-1}\cong \Delta_B + \Delta_B g(\sigma F) \Delta_B, ~~ \Delta_B
\equiv (m^2_1 - D^2_\mu-e_1 \sigma B)^{-1}\label{36}\ee

\be (m_2^2 - \hat D^2)^{-1}\cong \bar\Delta_B - \bar\Delta_B g(\sigma F)
\bar\Delta_B, ~~ \bar\Delta_B = (m^2_2- \bar D^2_\mu-e_2 \sigma
B)^{-1}\label{37}\ee and (\ref{30a}) can be rewritten keeping only one
spin-spin term   as\be G_{q\bar q} (x,y) = G_{q\bar q}^{(0)} (x,y) - \lan tr
[\Gamma(m_1-i\hat p_1)  \Lambda_B \bar   g(\sigma_1F) \Delta_B \Gamma(m_2-i\hat
p_2) \bar \Delta_B g (\sigma_2 F)\bar \Delta_B]\ran_A.\label{38}\ee
 one readily obtains for $ \Gamma=\gamma_5$ \be \mathcal{M} \equiv tr [\gamma_5  (m_1- i
\hat p_1) (\sigma_1 F) \gamma_5 (m_2- i \hat p_2) (\sigma_2 F)]= tr [ (m_1- i
\hat p_1) (\sigma_1 F) ( (m_2- i \hat p_2) (\sigma_2 F))^T],\label{39}\ee

\be (m_1- i \hat p_1) (\sigma F_1)=\left(\begin{array} {ll}(m_1+\omega_1)
\sigma H_1 -i \vesig \vep  \vesig \veE_1,&(m_1+\omega_1) \sigma  \veE_1 -i
\vesig \vep  \vesig \veH_1\\
i \vesig \vep \vesig \veH_1+(m_1-\omega_1)  \vesig \veE_1,&i \vesig \vep \vesig
\veE_1+(m_1-\omega_1) \sigma  \veH_1 \end{array}\right),\label{40}\ee

\be ((m_2- i \hat p_2) (\sigma F_2))^T=\left(\begin{array} {ll} i \vesig \vep
\vesig \veE_2+(m_2+\omega_2)  \vesig \veH_2,&i \vesig \vep \vesig
\veH_2+(m_2+\omega_2) \vesig  \veE_2\\
 (m_2-\omega_2) \vesig \veE_2 -i \vesig \vep
\vesig \veH_2,&(m_2-\omega_2) \vesig \veH_2 -i \vesig \vep  \vesig
\veE_2\end{array}\right).\label{41}\ee

Combining (\ref{40}) and (\ref{41}), one obtains $(\veH_i \equiv \veH (x_i),
i=1,2)$.
$$\mathcal{M}= tr_\sigma \left\{ \vesig\veH_1\vesig \veH_2 (2m_1m_2+
2\omega_1\omega_2) - 2 (\vesig \veH_1)\vesig \vep (\vesig \veH_2)\vesig
\vep)+\right.$$

$$+\vesig \veE_1\vesig \veE_2(2m_1m_2- 2\omega_1\omega_2) + 2
(\vesig \veE_1)\vesig \vep (\vesig \veE_2)\vesig \vep)+$$\be+ \vesig
\veE_1\vesig \veH_2(-2 i\vesig \vep (\omega_1+\omega_2))+\left.
\vesig\veH_1\vesig \veE_2(2 i\vesig \vep
(\omega_1+\omega_2))\right\}\label{42}\ee

In what follows we disregard first the terms, containing $(\vesig \vep)$ or
$(\vesig \vep)(\vesig \vep)$. In the nonrelativistic limit, $|\vep| \to 0,
\omega_i\to m_i, $ one has \be \lan \mathcal{M}\ran_A= 8 m_1m_2 \lan H_i (x_1)
H_i(x_2) \ran_A.\label{43}\ee Field correlators are expressed via two scalar
correlators $D(x_1-x_2)$ and $D_1 (x_1-x_2)$ as  in (\ref{HE0}),

Comparing with the standard definition for the nonrelativistic hyperfine (hf)
term, one has  \be  V_{hf} =\frac{\vesig^{(1)}\vesig^{(2)}}{12m_1m_2} V_4^{(H)}
(r), ~~ V_4^{(H)} (r) = \int^\infty_{-\infty} d\nu \frac{g^2}{N_c} \lan H_i (x)
H_i (y)\ran,~~ \nu\equiv x_4-y_4,\label{85a}\ee where  \be u_\mu = x_\mu
-y_\mu, ~~ \mu=1,2,3,4;~~ r = |\vex-\vey|=|\veu|,~~\nu\equiv u_4.\label{86a}\ee

One can separate perturbative part in $D_1$

\be D_1 (x) = D_1^{\rm pert} (x) + D_1^{\rm np} (x), ~~ D_1^{pert}(x) =
\frac{16\alpha_s}{3\pi x^4} +O(\alpha^2_s),\label{47}\ee and define the
potentials $$  V_4^{(H)}  (r) = \int^\infty_{-\infty} d\nu \frac{g^2}{N_c} \lan
H_i (x) H_i(y)\ran= $$\be= \int^\infty_{-\infty} d\nu \left(3D (r,\nu) + 3D_1
(r,\nu) + 2 r^2 \frac{\partial D_1 (r,\nu)}{\partial r^2}\right)=V_4^{(D)} (r)
+ V_4^{(1)} (r),\label{48}\ee \be V_4^{(E)} (r) =\int^\infty_{-\infty} d\nu
\left(3D (r,\nu) + 3D_1 (r,\nu) + (3\nu^2+r^2) \frac{\partial D_1
(r,\nu)}{\partial r^2}\right)=V^{(D)}_4-V_4^{(1)} (r).\label{49}\ee

As it is known \cite{18**}, the nonperturbative part of $D_1 $ and $D $ yield
much lower input in $V_4^{(E,H)}$, the leading part is due to $D^{\rm pert}_1$,
i.e. $V_4^{(1)}$. Inserting this, one obtains \be V_4^{(H)}\equiv V_4^{(H) {\rm
pert}}(r) = \frac{32 \pi \alpha_s}{3} \delta^{(3)}(\ver),\label{50}\ee and the
result (\ref{85a}), (\ref{50}) coincides with the known nonrelativistic limit.

Now we turn to the relativistic case, $\omega_i \gg m_i$. First of all we note,
that

\be V_4^{(E)} \equiv V_4^{(E) {\rm pert}}(r) =  \frac{-32 \pi \alpha_s}{3}
\delta^{(3)}(\ver).\label{51}\ee

Looking at (\ref{42}), one can see that in the relativistic case, when
$\omega_i \gg m_i$, there is a cancellation in the spin-spin interaction,
 in the combination $$ 2(m_1m_2+\omega_1\omega_2) V_4^{(H){\rm pert}} +2(
m_1m_2-\omega_1\omega_2) V_4^{(E){\rm pert}}=$$ \be = 2\omega_1 \omega_2\cdot
2V_4^{(1)} + 2m_1m_2 2V_4^{(D)},\label{52}\ee and multiplying  this result with
$\Delta_B$, $\bar \Delta _B$  as in (\ref{38}), in the $B=0$ case we obtain for
$\Delta_B \approx \frac{1}{2\Omega_1^2},~~\bar\Delta_B =
\frac{1}{2\Omega_2^2},$

\be V_{hf} = \frac{\vesig^{(1)} \vesig^{(2)}}{12 \bar \omega_1 \bar\omega_2}
\left\{ V_4^{(H) }(r) \left(1+ \frac{\vep^2}{3\omega^2}\right) +
\frac{m^2}{\omega^2} V_4^D (r)\right\}.\label{93a}\ee

Here $\bar \omega_i = \frac{\Omega^2_i}{\omega_i},$ and $\Omega_i$ is defined
in    (\ref{52*})  one can derive that $\bar \omega_i \geq \omega_i^{(0)}$,
e.g. in the nonrelativistic limit $\omega_i^{(0)} \to m_i, ~~ \Omega_i \to m_i$
and also $\bar \omega_i \to m_i$. The same happens in relativistic case, when
$eB\gg \sigma$.  In what follows for $\bar \omega_1 \neq\bar \omega_2$ it is
implied, e.g., that $\bar \omega^2_{+-} = \bar \omega_1(+-)\bar \omega_2(+-)$.


We now turn to the case of nonzero $B$ and now take into account the
noncommutative $(2\times 2)$ terms in $H_\sigma$ and  in $V_{hf}$, which we
write in the total mass as \be M= \bar M - \mu_1 \sigma_{1z} + \mu_2
\sigma_{2z} + a \vesig_1 \vesig_2, \label{94}\ee where $\mu_i
=\frac{eB}{2\omega}, a=\frac{1}{12\omega^2} \lan V_4^H\ran$.

For  $\pi^0, \rho^0 (s_z=0)$ states one obtains a standard mixing of $\lan +-|$
and $\lan -+|$ states of $\lan \sigma_{1z}, \sigma_{2z}|$, with $V_{hf}$, where
now for $B\neq 0$ we distinguish \be a_{11}=\lan +- |a\vesig_1\vesig_2| +-\ran
= \frac{1}{12 \bar\omega^2_{+-}} \lan V_4^{(H)}\ran,\label{95a}\ee

\be a_{22}=\lan -+  |a\vesig_1\vesig_2| -+ \ran = \frac{1}{12 \bar\omega^2_{-+
}} \lan V_4^{(H)}\ran,\label{96a}\ee

\be 2a_{12}= 2a_{21} = \lan -+  |a\vesig_1\vesig_2| +-\ran = \frac{2}{12\bar
\omega_{-+}\bar \omega_{+-}} \lan V_4^{ H }\ran,\label{97a}\ee

We also define $M_{11}, M_{22}$ as follows \be M_{11} = (\bar M- (\mu_1 +
\mu_2)- a_{11}))_{\omega_{+-}}; ~~M_{22} = (\bar M+ (\mu_1 + \mu_2)-
a_{22}))_{\omega_{-+}}.\label{98a}\ee

Finally, from $\det (M-E)=0$ , we obtain the  eigenvalues of $M$, \be E_{1,2}
=\frac12 (M_{11} + M_{22} ) \pm \sqrt{\left( \frac{M_{22} -
M_{11}}{2}\right)^2+ 4a_{12}a_{21}}.\label{99}\ee

Here $\omega_{+-}$ and $\omega_{-+}$ correspond to the diagonal states of
$\Delta_B \bar \Delta_B$ , i.e. for neutral mesons  to the stationary values of
$\omega^{(0)}$ in the states with spin projections $\lan +-|$ and $\lan -+|$
respectively. It is clear, that at large $eB$, the values of $\omega$ behave
differently, i.e. $\omega_{+-} \sim  const$, while $\omega_{-+} \sim
\sqrt{eB}$, and $M_{22} \sim \sqrt{eB}$, hence at large $eB$ the nondiagonal
part of $M$ in (\ref{99}) is decreasing, and $M$ tends to its diagonal
eigenvalues. \be E_1 (eB \to \infty) \to M_{11}, ~~ E_2 (eB \to \infty) \to
M_{22}.\label{100a}\ee One can notice, that for the $\lan +-|$ state $M_{11}$
contains at large $eB$ the fast decreasing part, $ -a_{11} \sim  -\psi^2(0)$,
and the latter is large in modulus, $\psi^2(0) \sim   \sim eB$, hence leading
to the negative mass $E_1$.

This happens already for only colormagnetic contribution $\lan H_i H_k\ran$ to
$V_4^{(H)}$, while our conclusion in section 2  was, that eigenvalues of
 $(m^2_i-\hat D^2_i)$ are positive, together with the total mass eigenvalues of
the operator $\lan \frac{1}{m^2_2 - \hat D^2_1} \frac{1}{ m^2_2- \hat
D^2_2}\ran$. Hence we conclude, that the perturbation theory in $V_{hf}$ breaks
down at large $eB$ and one has  to replace  it with some modified form.
 However, as was understood already in \cite{2,3} even at $B=0$ the
 perturbation theory with the potential $V_{ss}^{(0)} (r) \sim c \delta^{(3)}
 (\ver)$ is diverging  since $V_{ss}^{(0)}$ for any $c<0$ ensures infinite number of bound states, which are
 physically irrelevant. Therefore one should in any case take into account the relativistic smearing of the hf interaction, which appears due to the time
  integration in (\ref{34a}), which is taken in (\ref{34a}) along the straight line, instead of the complicated relativistic trajectory of the quark with time  fluctuations, see \cite{14}.
  The resulting smearing length $\lambda\ga \lambda_{conf}$, where $\lambda_{\rm conf}$  is the scale of $D(x)$., connected to the gluelump mass \cite{35}, $\lambda_{\rm conf}
  \approx 0.1 \div 0.15$ fm. On the lattice $\lambda\geq a$, $a$ is the lattice unit, $a\approx 0.1 \div 0.24$ fm.    therefore we
   replace
 $V_{ss}^{(0)}(r)$ by a smeared out version,  e.g.

\be \tilde \delta^{(3)} (r) =\left( \frac{\lambda}{\sqrt{\pi}}\right)^3
e^{-\mu^2 r^2} ;~~\tilde V_{ss} (r) = c\tilde \delta^{(3)} (\ver), ~~
\mu=\frac{1}{\lambda} \cong (1  \div 2)~{\rm GeV}. \label{101}\ee Using the
wave function $\tilde \psi(\eta)$ from (\ref{q53}) one obtains for $\lan \tilde
V_{ss}\ran$, in the PS meson \be \lan \tilde V_{ss} \ran =
\frac{c\mu^3}{\pi^{3/2} \sqrt{1+ \mu^2 r^2_z} (1+ \mu^2 r^2_\bot)},
c=-\frac{8\pi \alpha_s}{3\omega_1 \omega_2}.\label{103}\ee

For $\mu\to \infty$ one regains the original answer, $ \lan \tilde V_{ss}\ran
\to c \psi^2(0)$.

Since $r^2_\bot = \frac{2}{eB} \left( 1 + \frac{(\bar c\sigma)^2}{(eB)^2}
\right)^{-1/2}, r_\bot (eB \to \infty) \to 0, $ and $\lan \tilde V_{ss}\ran$
tends to a constant limit at large $eB$, preventing in this way breakdown of
the vacuum due to vanishing of the meson mass.

\section{The tensor forces in magnetic field}

As was established in  the previous section, the hf interaction, has the  form
(\ref{93a}) and in the m.f. the coefficients $\frac{1}{\bar
\omega_1\bar\omega_2}$ transform into $\frac{1}{\bar \omega_{ik}
\bar\omega_{i'k'}}$ where $(ik), (i'k')= (\sigma_{1z}, \sigma_{2z})= (+-)$ or
$(-+)$, for $S_{z}=0$  see Eqs. (\ref{95a})-(\ref{97a}). We now turn to the
tensor forces and discuss, how they are  transformed  in the m.f. As was found
in \cite{15,18} the total spin-dependent forces without m.f. can be written as
\be V_3^{ss} + V_4^{ss} \equiv \frac{\vesig^{(1)} \vesig^{(2)}}{12\omega_1
\omega_2} V_4 (r) + \frac{1}{12\omega_1 \omega_2} (3 (\vesig^{(1)}
\ven)(\vesig^{(2)} \ven)-\vesig^{(1)} \vesig^{(2)}) V_3(r)\label{5.1}\ee

It is clear from (\ref{5.1}), that for the spherically symmetric $q\bar q$
states the term $V_3(r)$ is irrelevant, and therefore requires a nonzero
angular momentum.

The situation is drastically changing in the m.f., since in this case the  form
of the wave function is distorted as in the elongated ellipsoid.

This fact implies the appearance of tensor forces in m.f. even in the ground
$q\bar q$ state with the zero angular momentum. This situation was studied for
the hydrogen atom case in m.f. in \cite{34} in the nonrelativistic treatment of
the tensor  forces in the $q\bar q)$ system.

We are using again the correlator technic and the perturbative correlator
$D^{\rm pert}_1$, in terms of which the tensor interaction can be written as in
(\ref{33a}).

The expectation value of the tensor term in the ground state with the wave
function $\tilde\psi(\rho, z )$ (\ref{q53}) can be written as

\be \lan V_3^{ss}\ran =  \frac{8 \alpha_s \sigma_i^{(1)} \sigma^{(2)}_k}{9 \pi
\omega_i\omega_2} \int^\infty_{-\infty} d\nu \rho d\rho dz d \varphi (3n_i n_k-
\delta_{ik}) \psi^2 (\rho, z) (-r^2) \frac{\partial}{\partial r^2}
\frac{1}{(r^2+ \nu^2)^2}\label{5.3}\ee Using the relations

\be \frac{1}{(r^2+\nu^2)^3} = \frac12 \left. \frac{\partial^2}{\partial
\alpha^2}\right|_{\alpha=0} \frac{1}{r^2+\nu^2 +\alpha}; ~~ \frac{1}{r^2 +\nu^2
+\alpha} = \int^\infty_0 d\beta e^{-\beta (r^2 + \nu^2+ \alpha)}.\label{5.4}\ee
 and the $q\bar q$ wave function in the zeroth approximation

 \be \psi^2 (\rho, z) = N  e^{-\frac{\rho^2}{r_\bot^2}-
\frac{z^2}{r^2_3}};  ~~ N= \frac{1}{\pi^{3/2} r^2_\bot r^2_3},\label{5.5}\ee
one can calculate the sum

\be \lan V_3^{ss}\ran+ \lan V^{ss}_4\ran = \frac{8\pi \alpha_s}{9 \omega_1
\omega_2}   N \left\{ \vesig_1 \vesig_2 + \frac{K}{4} \left(
\frac{1}{r^2_\bot}-\frac{1}{r^2_3} \right) (-\sigma_x^{(1)}
\sigma_x^{(2)}-\sigma_y^{(1)} \sigma_y^{(2)}+ 2 \sigma_z^{(1)}
\sigma_z^{(2)})\right\}, \label{5.6}\ee
 Here $K$ is
\be K= r^2_\bot \int^\infty_0 \frac{2u^{4} du}{(u^2+1)^2
(u^2+\frac{r^2_\bot}{r^2_3})^{3/2}}.\label{5.7}\ee

Note, that we have omitted here for simplicity in $\lan V_4\ran^{ss} \sim
\vesig_1 \vesig_2$, the additional terms, which appear in (\ref{93a}). The
integral $K$ in (\ref{5.7}) can be done explicitly, using the relation \be
\int^\infty_0 \frac{du}{(u^2+p^2) (u^2+q^2)^{1/2}} = \frac{1}{2p\sqrt{p^2-q^2}}
\ln \left( \frac{p+ \sqrt{p^2-q^2}}{p-\sqrt{p^2-q^2}}\right).\label{5.8}\ee

It is important, that $K\leq r^2_\bot$, therefore since $r^2_\bot\to 0 $ for
$eB \to \infty$, see eq. (\ref{q54}), the relative role of the tensor term in
(\ref{5.6}) is diminishing with growing $eB$, and the absolute  value of $\lan
V_3^{ss}\ran $ depends only on the values of $\omega_1, \omega_2$.
 To find these values, we again as in (\ref{94}) write the total mass, but now
 with the addition of the  tensor term and for any quark charges
 \be M=\bar M + a\vesig_1\vesig_2+ c^{(1)} \sigma_{1z} + c^{(2)}_{2z}+ b( - \sigma_x^{(1)}
 \sigma_x^{(2)}-\sigma_y^{(1)}\sigma_y^{(2)}+2\sigma_z^{(1)}\sigma_z^{(2)})\equiv
 \bar M + h,\label{5.9}\ee
where \be a= \frac{8\pi\alpha_s}{9\bar\omega_1\bar\omega_2} N, ~~ c^{(1)} = -
\frac{e_1 B}{2\omega_1},~~c^{(2)} = - \frac{e_2 B}{2\omega_2},\label{5.10}\ee

\be b= \frac{8\pi\alpha_s}{9\bar\omega_1\bar\omega_2} N \frac{K}{4} \left(
\frac{1}{r^2_\bot}-\frac{1}{r^2_3} \right) .\label{5.11}\ee Moreover, one must
distinguish the values of coefficients $a,b,c_i$ in different spin projection
states $(\sigma_{1z} \sigma_{2z})$ namely the values of $\omega_i$ in $a,b,c_i$
in different $(\sigma_z^{(1)}, \sigma_z^{(2)})$ states. E.g. for $S_z=+1 (-1)$
one must write $\bar\omega_1\bar\omega_2 \to \bar\omega^2_{++}
(\bar\omega^2_{--})$, while for $S_z=0$, one has  the  matrix elements as in
(\ref{95a})-(\ref{97a}). In a similar way for our general case with tensor
forces in (\ref{5.6}), one can write

 \be \begin{array}{ll} \lan +-
|h| +-\ran =& -a_{11} - 2b_{11} +c_{11}^{(1)}- c_{11}^{(2)},\\
\lan -+
|h| -+\ran =& -a_{22} - 2b_{22} +c_{22}^{(1)}- c_{22}^{(2)},\\
\lan +- |h| -+\ran =& \lan -+ |h|+-\ran =  2(a_{12}-b_{12}) =
2(a_{21}-b_{21}),\end{array}\label{5.12}\ee and all diagonal terms have the
corresponding $ \bar\omega_i$, e.g. $ \bar\omega_1= \bar\omega_2 =
\bar\omega_\pm$ for $a_{11}, b_{11}, c_{11}^{(i)},$ and in nondiagonal matrix
elements $ \bar\omega_1 \bar \omega_2 =  \bar\omega_{+-} \bar\omega_{-+}$.

From $\det(h-E) =0$ with elements in (\ref{5.12}) one obtains two eigenvalues
of the spin-dependent part $E_1$, $E_2$.

The resulting expressions for $E_1, E_2$ coincide with  those in (\ref{99}),
when one makes the following replacements \be  a_{11} \to a_{11} + 2b_{11}, ~~
a_{22} \to a_{22} + 2b_{22}, ~~ a_{12 }\to a_{12}- b_{12},\label{5.14}\ee and
$\mu_1 \to -c^{(1)}$.

Eqs. (\ref{q51}), (\ref{q52}) contain  a prescription for the values of
$\omega_{ij}$, $i,j =+,-,$ entering in $M_{11}, M_{22}$, which is valid also in
the case of nonzero tensor forces.

\section{Conclusions and prospectives}

We have obtained explicit relativistic expressions for the spin-spin
interaction terms in the $q\bar q$ system in the arbitrarily strong m.f. As a
by-product we also obtained in the same formalism expressions for the np
self-energy corrections in m.f. These formulas are generalizations of the
previously found expressions for the $q\bar q$ mesons in absence of  m.f., see
e.g. \cite{15}, and we also found corrections to those expressions, see Eq.
(\ref{93a}), where the terms $\frac{\vep^2}{3\omega^2} V_4^{(H)} (r)$ and
$\frac{m^2}{\omega^2} V_4^{(D)}(r)$ are new.

It is remarkable, how m.f. changes the spin-spin forces. First of all, the
matrix element of the hf term $V_4^{(H)} (r) \sim \delta^{(3)}(\ver)$, in the
strong m.f. is proportional to the $\psi^2(0)$ -- the probability of coming
together of $q$ and $\bar q$, which in strong m.f. grows as $eB$

This effect is known in nonrelativistic case, where the hf term is $\lan
V_{hf}\ran \sim \frac{\psi^2(0)}{M_1M_2} \sim \frac{eB}{M_1M_2}$ and was
discussed recently in \cite{34} for the hydrogen  case. However, in this case
$M_2 =M_p$ is very large and $\lan V_{hf} (hydr)\ran $ is small and this growth
of $\lan V_{hf}\ran$ was thought  to cease in relativistic limit, $eB \sim
m^2_e$.

However, as we have shown here in the paper, the growth of $\lan V_{hf}\ran$ in
the relativistic case for $eB\to \infty$ is possible whenever $\omega^2$ in the
denominator of $\lan V_{hf}\ran \sim\frac{\psi^2(0)}{\omega^2}$ does not grow
with $eB$, which occurs for the $q\bar q$ states,  where magnetic moment terms
compensate the growth of the rest part of mass, e.g. for the $<+-|$ states of
the neutral mesons, like $\pi^0$. Then hf terms yield negative $(-3\lan
V_{hf}\ran)$ contribution to the mass, linearly growing with $eB$ in modulus,
thus giving the absurd negative mass result.

To disprove this result, we have given in Section 2  the proof, that  the term
$V_4^{(H)}$, which causes this problem, cannot generate negative mass in any
m.f., and we are coming to the conclusion, that this discrepancy occurs due to
the use of the perturbation theory for the hf term, proportional to the
delta-function. Therefore we have suggested in section 4 the smearing procedure
of the hf term, which takes into account  the relativistic Zitterbewegung of
quarks and should be supplemented with a  rigorous procedure of the summation
or estimation  of the whole perturbative series. We stress, that this situation
occurs solely due to m.f., which cannot, as shown in section 2, provide the
pair creation and vacuum reconstruction, in contrast to the real Minkowskian
electric fields, which are capable of the pair creation (as in the Schwinger
famous formula \cite{20}) and of the vacuum reconstruction with emission of
positrons (as in superheavy atoms with $Z>Z_{\rm crit}).$

Another interesting new effect is the appearance  of the tensor force effects
at nonzero m.f., which can be tested both in atoms and mesons. However, as we
have  shown above, at large m.f. the tensor force contribution does not  grow
with $eB$, unlike hf term, and is always smaller than the latter.

In this way we have completed the main part of the strong dynamics of mesons in
magnetic field for the zero angular momentum, started in \cite{14}.

The author  is grateful for useful discussions to M.A.Andreichikov,
S.I.Godunov,  B.O.Kerbikov, V.D.Orlovsky,  and  M.I.Vy\-sotsky  and to the
members of theoretical seminar at FIAN (Moscow).



\end{document}